# Designing Asymmetric Multiferroics with Strong Magnetoelectric Coupling

X. Z. Lu and H. J. Xiang*

Key Laboratory of Computational Physical Sciences (Ministry of Education), State Key Laboratory of Surface Physics, and Department of Physics, Fudan University, Shanghai 200433, P. R. China

e-mail: hxiang@fudan.edu.cn

**Abstract**

Multiferroics offer exciting opportunities for electric-field control of magnetism. Unfortunately, single-phase multiferroics suitable for such applications at room temperature has not been discovered. Here, we propose the concept of a new type of multiferroics, namely, "asymmetric multiferroic". In asymmetric multiferroics, two locally stable ferroelectric states are not symmetrically equivalent, leading to different magnetic properties between these two states. Furthermore, we predict from first-principles that a Fe-Cr-Mo superlattice with the $LiNbO_3$-type structure is such an asymmetric multiferroic. The strong ferrimagnetism, high ferroelectric polarization, and significant dependence of the magnetic transition temperature on polarization make this asymmetric multiferroic an ideal candidate for realizing electric-field control of magnetism at room temperature. Our study suggests that asymmetric multiferroic may provide a new playground for voltage control of magnetism and find its applications in spintronics and quantum computing.

In recent years, multiferroics, displaying magnetic, polar, and elastic order parameters simultaneously, have attracted numerous research interests [1-6]. Among them, room-temperature multiferroics with strongly coupled large ferromagnetism and ferroelectricity are particularly intriguing because they offer a route to achieve the electric-field control of magnetism in realistic information storage and processing applications. $BiFeO_3$ is a well-known room temperature multiferroic material, but it is antiferromagnetic (AFM) [7]. Although Z-type hexaferrites were discovered to exhibit a low-field magnetoelectric (ME) effect at room temperature, the spin-order induced electric polarization is rather tiny (about 20 $\mu C/m^2$) [8]. On the other hand, for all current known ferromagnetic (FM)-ferroelectric (FE) multiferroics [9-12], the Curie temperatures are far below room temperature and there is no intrinsic coupling between the FM and FE orders. Therefore, search for room temperature multiferroics suitable for realistic applications may require conceptual breakthrough.

In this work, we propose the concept of "asymmetric multiferroic" as a new type of multiferroics. The symmetry breaking between the two locally stable FE states of an asymmetric multiferroic may result in a strong dependence of the magnetic properties on the changing of FE polarization. Furthermore, a general strategy based on compositional inversion symmetry breaking [13] and spin-lattice coupling is suggested to design asymmetric multiferroics. By applying this method to $LiNbO_3$-type structures, we predict an asymmetric multiferroic oxide superlattice (i.e., $LiNbO_3$-type $Al_3Y_3Fe_3MoCr_2O_{18}$) in which the phase competition between ferrimagnetic and paramagnetic states is determined by the direction of the polarization. Thus, the room-temperature electric field control of strong ferrimagnetism could be realized in this system.

In usual single-phase multiferroics, such as BiFeO$_3$ and TbMnO$_3$, two locally stable FE states are symmetrically equivalent since they are related to each other by a symmetry operation ($\hat{R}$) of point group (e.g. inversion symmetry in most cases). For this reason, we call these systems "symmetric multiferroics", as shown in Fig. 1(a). It is clear that the magnetic properties of these states are also symmetrically equivalent. As already mentioned above, it is very difficult to identify or design room temperature symmetric multiferroics with strong coupled FE-FM orders. Here, we propose the existence of another type of multiferroics, namely, "asymmetric multiferroics" [see Fig. 1(b)]. Similar to symmetric multiferroics, there exist two locally stable states with different FE polarizations. However, these two states are not equivalent since they are not related to each other by any symmetry operation of point group. Thus, the magnetic properties (e.g., net magnetization, magnetic transition temperature, magnetic anisotropy) between these two states may be dramatically different. When an electric field induces the transition between the two states, it also changes the magnetic properties. This intrinsic ME coupling may make asymmetric multiferroics particularly appealing for realizing electric-field control of magnetism. Another difference between the two types of the multiferroics is that the transition state (TS) between the two FE states has a non-polar structure (i.e. paraelectric state) in symmetric multiferroic, whereas it has a polar structure in asymmetric multiferroic (see Fig. 1). To design a useful asymmetric multiferroic, there are two main issues to be resolved. First, how to realize two locally stable FE states? Second, how to make the magnetic properties between the two states as different as possible? To this end, we will demonstrate below a general design principle based on compositional inversion symmetry breaking [13] and strong spin-lattice coupling.

We will take LiNbO$_3$-type structure [see Figs. 2(a) and (b)] as an example to design asymmetric multiferroics since LiNbO$_3$-type systems were found to display robust polar distortions [14,15]. Our recent first-principles study [16] revealed that the polar instability in most LiNbO$_3$-type systems is commonly driven by A-site ion due to the tendency of A-site ion to form ionic or covalent bonds with the neighboring out-of-plane O ions. The polar instability of LiNbO$_3$-type ABO$_3$ system can be qualitatively estimated by the tolerance factor defined for the $R\bar{3}c$ LiNbO$_3$-type structure $t_R = \frac{r_A + r_O}{\sqrt{2}(r_B + r_O)}$, where $r_A$, $r_B$ and $r_O$ are the ionic radii [17] of the A-site ion, B-site ion and O ion, respectively. When $t_R < 1$ as found in most systems, the A-site ion is too far from the out-of-plane O ion, thus there will be an A-site driven polar instability.

We find that LiNbO$_3$-type compound ABO$_3$ with $t_R < 1$ is FE even when the B-site ion (e.g. Cr ion) is magnetic (see part 2 of [18]). This is important because the ferroelectricity is usually incompatible with magnetism in most cases [2]. While ACrO$_3$ (a symmetric multiferroic according to our definition) is both FE and magnetic (FM or AFM), there is almost no coupling between the FE order and magnetic order. We find that the intrinsic coupling between the polar order and magnetic order can be achieved by tuning the composition of the ABO$_3$ structure. As shown in Fig. 2(c), the TSs of ABO$_3$, A$_2$BB′O$_6$, AA′B$_2$O$_6$ systems based on the $R\bar{3}c$ structure are non-polar (i.e. paraelectric state) and the two FE states (**P** and –**P** states) evolved from the TS are symmetrically equivalent. Thus, they belong to symmetric multiferroics. On the contrary, the TS of AA′BB′O$_6$ based on the $R\bar{3}c$ structure is polar ($R3$ space group). This means that AA′BB′O$_6$ can be used to design an asymmetric multiferroic. We then test this idea by considering AlYCrWO$_6$ as an example. Our calculations show that AlYCrWO$_6$ with **P** along the hexagonal c-axis (P$_{up}$ state) has an insulating AFM ground state, while the FE state with **P'** along the negative hexagonal c-axis ($P_{dn}$ state) has an insulating FM ground state. This is very interesting since an applied electric field along the c axis could switch the $P_{dn}$ state to the $P_{up}$ state, thus change the magnetic ground state from the FM state to AFM state. It should be noted that this intrinsic ME coupling does not require spin-orbit coupling, thus could be very strong. Unfortunately, the critical temperatures [77 K and 151 K from Monte Carlo (MC) simulations] of the AFM and FM states are much lower than room temperature, which excludes its realistic applications.

To increase the critical temperature, we turn to the ferrimagnetic systems since the insulating ferrimagnetism might sustain at much higher temperature unlike the insulating ferromagnetism. We find that FE $A_2FeMoO_6$ (A = Al, Ga, In, Sc, Y) (a symmetric multiferroics) are ferrimagnetic with the Curie temperature above the room temperature. In these systems, both $Fe^{3+}$ and $Mo^{3+}$ ions adopt the high-spin configuration with the magnetic moments of 5 $\mu_B$ and 3 $\mu_B$, respectively. The presence of strong FE polarization and high spontaneous magnetization at the room temperature makes $A_2FeMoO_6$ promising for future multi-state memory applications. In order to realize the ME coupling, we consider the asymmetric multiferroic AA′BB′O$_6$ systems related to $A_2FeMoO_6$. Our calculations show that AlA′FeCrO$_6$ and AlA′FeMoO$_6$ (A′ = Al, Ga, In, Sc, Y) have large electric polarization which is even higher than that of $BaTiO_3$ (For instance, the difference in the electric polarization between two states with opposite polarizations are 114 and 120 μC/cm$^2$ for AlYFeCrO$_6$ and AlYFeMoO$_6$, respectively). They are also ferrimagnetic with a rather high Curie temperature. In particular, the Curie temperature $T_c$ for AlA′FeMoO$_6$ is much higher than room temperature, while $T_c$ for AlA′FeCrO$_6$ is around 200 K, slightly below room temperature. Interestingly, there is a large difference in the Curie temperature between the $P_{dn}$ and $P_{up}$ states for both AlA′FeCrO$_6$ and AlA′FeMoO$_6$ (A′≠Al). This temperature difference becomes larger along with the increase of the size difference between the A′ ion and Al ion since the degree of the compositional symmetry breaking depends on the ionic radius difference. For example, the difference in the Curie temperature between the $P_{dn}$ and $P_{up}$ states for $AlYFeMoO_6$ is higher than 300 K. The different magnetic properties between the $P_{dn}$ and $P_{up}$ states arise from the spin-lattice coupling effect. We illustrate this point by considering $AlYFeCrO_6$. As shown in the inset of Fig. 3(b), there are two different Fe-Cr symmetric exchange interactions ($J_1$ and $J_2$) in $AlYFeCrO_6$. The computed effective exchange interaction parameters are $J_1 = 12.90$ meV and $J_2 = 17.83$ meV ($J_1 = 5.64$ meV and $J_2 = 14.88$ meV) for the $P_{dn}$ ($P_{up}$) state. It can be seen that there is a large difference in $J_1$ between the $P_{dn}$ and $P_{up}$ states. From the density of state (see Figure S2 of [18]) and maximally localized Wannier function (MLWF) analysis [19], we find that the hybridization between the occupied majority-spin $t_{2g}$ orbitals of $Cr^{3+}$ and unoccupied minority-spin $e_g$ orbitals of $Fe^{3+}$ is mainly responsible for the AFM Fe-Cr interaction [see Fig. 3(a)]. The hopping parameters ($t_{\pi\sigma}$) will be different because the Fe-O and Cr-O (O is the corner-shared O ion) bond length changes dramatically between the $P_{dn}$ and $P_{up}$ states [see inset of Fig. 3(a)]. From the MLWF calculation, $t_{\pi\sigma}(P_{dn})$ (about 0.12 eV) is found to be larger than $t_{\pi\sigma}(P_{up})$ (about 0.09 eV) possibly because the Cr-O bond length in the case of $P_{dn}$ state is shorter so that the more localized $Cr^{3+}$ $t_{2g}$ orbitals can interact better with the more delocalized $Fe^{3+}$ $e_g$ orbitals. A stronger π-σ hopping leads to more energy gain of the AFM state, explaining why $J_1(P_{dn})$ is much larger than $J_1(P_{up})$. The mechanism for the ME coupling in $AlA′FeCrO_6$ and $AlA′FeMoO_6$ is similar to that in the $La_{0.7}Sr_{0.3}MnO_3/BiFeO_3$ interface [20], where the switch of FE polarization changes the exchange coupling between $La_{0.7}Sr_{0.3}MnO_3$ and $BiFeO_3$. The key difference is that the ME coupling in $AlA′FeCrO_6$ and $AlA′FeMoO_6$ is a bulk effect instead of an interfacial phenomenon.

The large difference in the Curie temperature between the $P_{dn}$ and $P_{up}$ states suggests that it is possible to tune the magnetism using an electric field. At a temperature between $T_c(P_{dn})$ and $T_c(P_{up})$, an electric field can control the switch between the $P_{dn}$ state with the ferrimagnetic order and the $P_{up}$ state with the paramagnetic order. As shown in Fig. 3(b), both $T_c(P_{dn})$ and $T_c(P_{up})$ are higher (lower) than the room temperature for AlA′FeMoO$_6$ (AlA′FeCrO$_6$). Thus, both systems are not suitable for room temperature applications. We find that it is possible to tune the Curie temperature close to the room temperature by appropriately mixing Cr and Mo in the superlattice. Namely, decreasing the Cr concentration will make the Curie temperature higher and vice versa. The superlattice (hereafter referred as Fe-Cr-Mo superlattice) shown in Fig. 4(b) is such an example. The MC simulations show that $T_c(P_{dn})$ is higher than the room temperature, while $T_c(P_{up})$ is lower than the room temperature. This suggests that it is possible to realize the room-temperature electric-field switch between the $P_{dn}$ state with the ferrimagnetic order (average $|\mathbf{M}| = 2.0$ μ$_B$ per Fe) and the $P_{up}$ state with the paramagnetic order ($|\mathbf{M}| = 0$) in such an asymmetric multiferroic [see Fig. 4(b)]. Calculations including the spin-orbit coupling effect indicate that the easy axis of the $P_{dn}$ state is along c-axis (anisotropic energy is -0.03 meV/Mo), while the ferrimagnetic state of the $P_{up}$ state has an in-plane anisotropy (anisotropic energy is -0.57 meV/Mo). The different magnetic anisotropy suggests that it is also possible to tune magnetization direction by an electric field at a temperature lower than $T_c(P_{up})$.

The asymmetric multiferroic Fe-Cr-Mo superlattice proposed in this work has strong electric polarization and high spontaneous magnetization, and both its FE and magnetic critical temperature are higher than room temperature. Importantly, the coupling between the polar order and magnetic order is intrinsic and strong. These excellent properties of the Fe-Cr-Mo superlattice make it an ideal candidate for realizing optimal electric-field control of magnetism, which should satisfy the following requirements in order to reduce power consumption and minimize device size: (1) voltage control instead of current control; (2) switching of a single bulk phase instead of one single interface; (3) non-volatile; (4) operation at room temperature. Despite numerous efforts [21-25] in realizing electric-field control of magnetism, the optimal electric-field control of magnetism has never been achieved so far. Our study may open up new avenue towards this end.

We note that some symmetric multiferroics based on the LiNbO$_3$-type structure have been proposed previously in the literatures. Hao *et al.* [26] predicted that LiNbO$_3$-type PbNiO$_3$ is a multiferroic with large electric polarization but zero net magnetization. Considering spin-orbit coupling effect, Fennie [27] proposed that a polar lattice distortion induces weak ferromagnetism in LiNbO$_3$-type materials such as MTiO$_3$ (M = Fe,Mn,Ni) and the direction of the weak magnetization may be reversed with an applied electric field assuming that the AFM order parameter does not change with the electric polarization. The prediction is partially validated experimentally [28] by confirming that FeTiO$_3$ is FE at room temperature and weakly ferromagnetic below 120 K. However, the magnetization reversal by electric field has not been achieved in this low-temperature multiferroic to our best knowledge.

Finally, we discuss the stability of the LiNbO$_3$-type superlattices and the possible synthesis route. As shown in Figure S3 of [18], phonon dispersions indicate that both AlYFeCrO$_6$ and AlYFeMoO$_6$ are dynamically stable. Thus, although these superlattices may not be the global lowest energy structures, they are locally stable. The energy barrier between the two states with opposite electric polarizations is also found to be sufficiently high so that the FE state is stable at least up to room temperature. For example, the energy barrier for AlYFeCrO$_6$ from the climb nudged elastic band calculation [29] is around 0.20 eV/O, which is higher than that (0.05 eV/O) in room-temperature FE LiNbO$_3$. Our first-principles molecular dynamics simulations suggest that the FE states are stable at least up to 500 K. LiNbO$_3$-type metastable phase such as FeTiO$_3$ with space group *R*3*c* has been successfully synthesized under high pressure and temperature [28]. Furthermore, LiNbO$_3$-type ZnSnO$_3$ thin film was epitaxially grown on a (111)-SrRuO$_3$/(111)-SrTiO$_3$ substrate by pulsed laser deposition [30]. We expect that the LiNbO$_3$-type superlattices proposed in this work may be synthesized by pulsed laser deposition or molecular beam epitaxy technique.

In summary, we propose the concept of a new type of multiferroics-"asymmetric multiferroic". Furthermore, we predict a new asymmetric multiferroic with the LiNbO$_3$-type structure (i.e., Al$_3$Y$_3$Fe$_3$MoCr$_2$O$_{18}$ superlattice). In this asymmetric multiferroic, there are large FE polarization and high spontaneous ferrimagnetic magnetization at room temperature. More importantly, the FE order is intrinsically coupled with the magnetic order, leading to a new route to electric-field control of magnetism in which the paramagnetic-ferrimagnetic phase transition at the room temperature can be induced by electric field. The main guiding principle for designing asymmetric multiferroics is to make the two FE states inequivalent by alloying and tune the magnetic properties by exploiting spin-lattice coupling, which can be broadly applied to other systems such as perovskite oxides. Our study suggests that asymmetric multiferroics open up a new avenue of electric-field control of magnetism at room temperature.

**References**

[1] S.-W. Cheong and M. Mostovoy, Nature Mater. **6**, 13 (2007).

[2] R. Ramesh and N. Spaldin, Nature Mater. **6**, 21 (2007); N. A. Hill, J. Phys. Chem. B **104**, 6694 (2000).

[3]  S. Picozzi and C. Ederer, J. Phys. Condens. Matter **21**, 303201 (2009).

[4]  K.Wang, J.-M. Liu, and Z. Ren, Adv. Phys. **58**, 321 (2009).

[5]  D. Khomskii, Physics **2**, 20 (2009).

[6]  H. J. Xiang, E. J. Kan, Y. Zhang, M.-H. Whangbo, and X. G. Gong, Phys. Rev. Lett. **107**, 157202 (2011); P. S. Wang and H. J. Xiang, Phys. Rev. X **4**, 011035 (2014).

[7]  I. Sosnowska, T. Peterlin-Neumaier, and E. Steichele, J. Phys. C: Solid State Phys. **15**, 4835 (1982).

[8]  Y. Kitagawa, Y. Hiraoka, T. Honda, T. Ishikura, H. Nakamura, and T. Kimura, Nature Mater. **9**, 797 (2010).

[9]  D. L. Janes, R. E. Bodnar, and A. L. Taylor, J. Appl. Phys. **49**, 1452 (1978).

[10] G. A. Smolenskiĭ and I. E. Chupis, Sov. Phys. Usp. **25**, 475 (1982).

[11] C. J. Fennie and K. M. Rabe, Phys. Rev. Lett. **97**, 267602 (2006); J. H. Lee *et al.*, Nature **466**, 954 (2010).

[12] J. H. Lee and K. M. Rabe, Phys. Rev. Lett. **104**, 207204 (2010).

[13] N. Sai, B. Meyer, and D. Vanderbilt, Phys. Rev. Lett. **84**, 5636 (2000).

[14] Y. Inaguma, M. Yoshida, and T. Katsumata, J. Am. Chem. Soc. **130**, 6704 (2008).

[15] Y. Shi *et al.*, Nat. Mater. **12**, 1024 (2013).

[16] H. J. Xiang, arXiv:1312.4225.

[17] R. D. Shannon, Acta Cryst A **32**, 751 (1976).

[18] See supporting materials for computational methods, results for $ACrO_3$, partial density of states of $AlYFeCrO_6$, phonon dispersions for $AlYFeCrO_6$ and $AlYFeMoO_6$.

[19] N. Marzari and D. Vanderbilt, Phys. Rev. B **56**, 12847 (1997); A. A. Mostofi *et al.*, Comput. Phys. Commun. **178**, 685 (2008).

[20] P. Yu, Y. H. Chu, and R. Ramesh, Phil. Trans. R. Soc. A **370**, 4856 (2012).

[21] H. Ohno *et al.*, Nature **408**, 944 (2000).

[22] T. Lottermoser *et al.*, Nature **430**, 541 (2004).

[23] M. Weisheit *et al.*, Science **315**, 349 (2007).

[24] V. Garcia *et al.*, Science **327**, 1106 (2010).

[25] Y. Yamada *et al.*, Science **332**, 1065 (2011).

[26] X. F. Hao, A. Stroppa, S. Picozzi, A. Filippetti, and C. Franchini, Phys. Rev. B **86**, 014116 (2012).

[27] C. J. Fennie, Phys. Rev. Lett. **100**, 167203 (2008).

[28] T. Varga *et al.*, Phys. Rev. Lett. **103**, 047601 (2009).

[29] G. Henkelman, B. P. Uberuaga, and H. Jónsson, J. Chem. Phys. **113**, 9901 (2000).

[30] J. Y. Son *et al.*, J. Am. Chem. Soc. **131**, 8386 (2009).

**Acknowledgements**

Work was supported by NSFC, FANEDD, NCET-10-0351, Research Program of Shanghai Municipality and MOE, the Special Funds for Major State Basic Research, Program for Professor of Special Appointment (Eastern Scholar), and Fok Ying Tung Education Foundation.

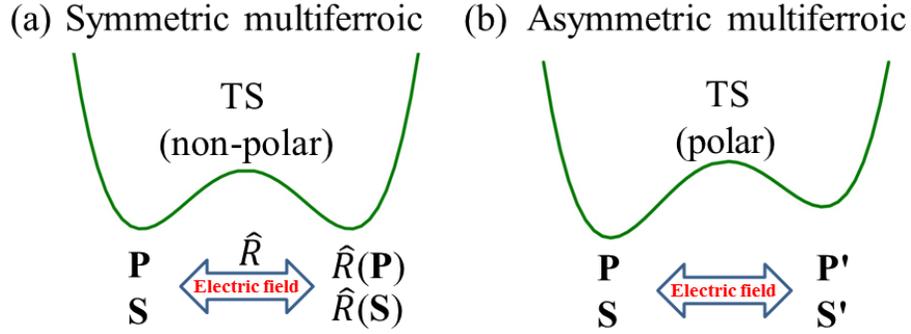

Figure 1. Schematic illustration of (a) symmetric multiferroic and (b) asymmetric multiferroic. TS, **P** and **S** represent the transition state, polarization and magnetic properties, respectively. $\hat{R}$ is the symmetry operation of point group. In a symmetric multiferroic, the two FE states and the associated magnetic properties are related by $\hat{R}$, and the TS is non-polar (i.e. paraelectric). On the contrary, in an asymmetric multiferroic, the two locally stable FE states switchable by electric field are not symmetrically equivalent and the TS is polar. Therefore, these two states have different magnetic properties.

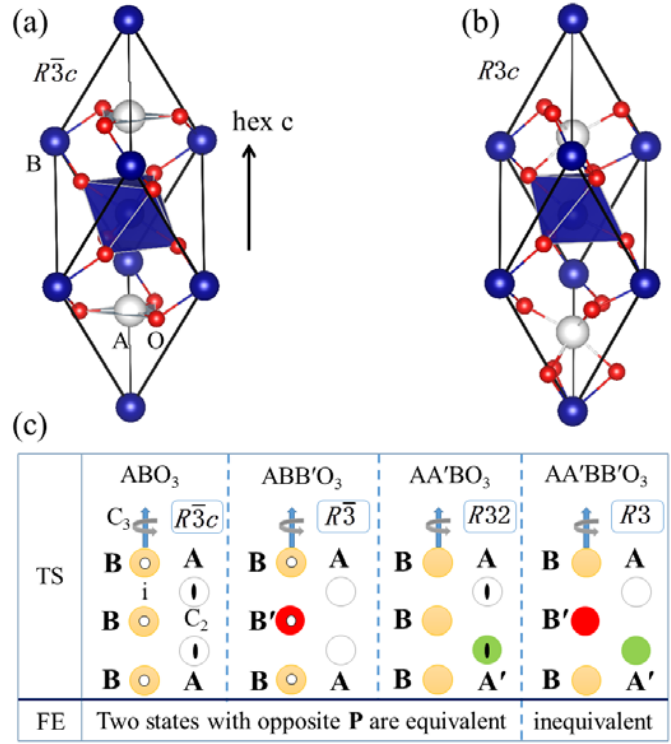

Figure 2. (a) and (b) show the paraelectric $R\bar{3}c$ and ferroelectric $R3c$ structures, respectively. The arrow indicates the hexagonal c axis. (c) illustrates how an asymmetric multiferroic can be designed by tuning the composition of the transition state (TS). The symmetry operations ($C_3$, $C_2$, i) are denoted. Only in the case of AA′BB′O$_6$, the two locally stable FE states evolved from the TS are inequivalent.

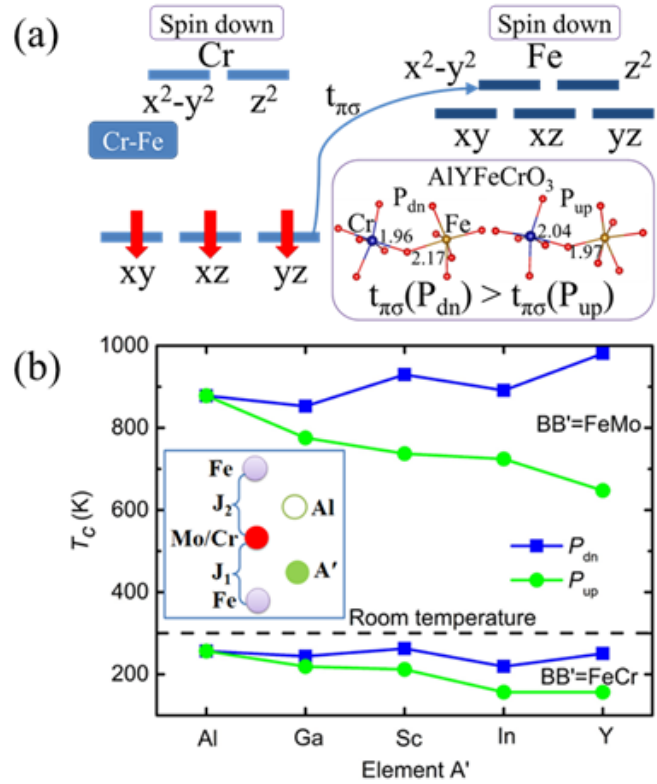

Figure 3. (a) illustrates that Cr-Fe $\pi$-$\sigma$ coupling is responsible for AFM coupling between a spin-down $Cr^{3+}$ ion and a spin-up $Fe^{3+}$ ion in $AlYFeCrO_6$. Inset: Comparison of the local structures and hopping parameters $t_{\pi\sigma}$ between two states with opposite polarizations ($P_{dn}$ and $P_{up}$) for the spin interaction $J_1$ [see inset of (b)]. (b) Curie temperatures of the $P_{dn}$ and $P_{up}$ states in $AlA'FeCrO_6$ and $AlA'FeMoO_6$ systems (A′ = Al, Ga, In, Sc, Y) from the MC simulations. Inset: Illustration of the superlattice and two different nearest neighboring spin exchange interactions ($J_1$ and $J_2$).

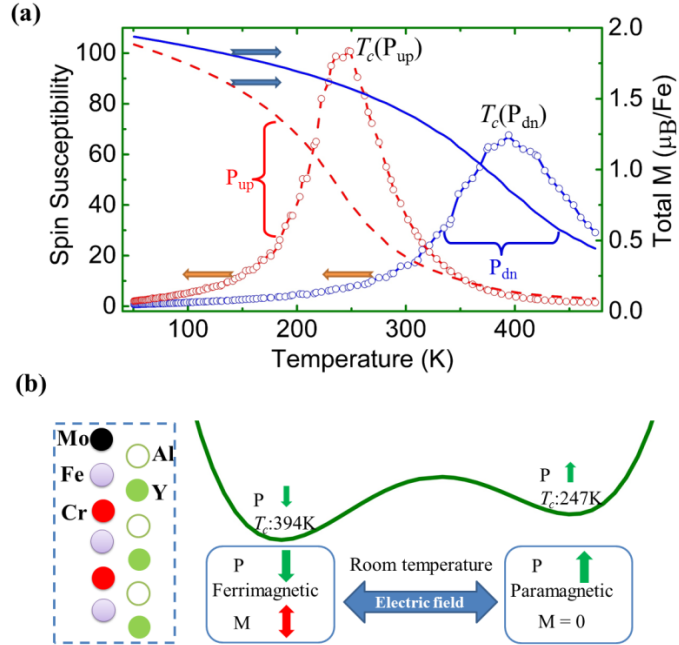

Figure 4. (a) Spin susceptibility and total magnetization versus temperature in both $P_{dn}$ and $P_{up}$ states for the Fe-Cr-Mo superlattice shown in (b). It shows that $T_c(P_{dn}) = 394$ K is higher than the room temperature, while $T_c(P_{up}) = 247$ K is lower than the room temperature. (b) Left panel: Schematic plot of the Fe-Cr-Mo superlattice (i.e., $Al_3Y_3Fe_3MoCr_2O_{18}$ structure); Right panel: Schematic plot of $P_{dn}$ and $P_{up}$ states for this system and electric-field induced phase transition from ferrimagnetic state with $P_{dn}$ to paramagnetic state with $P_{up}$ at the room temperature.